\documentclass{aa}  

\usepackage{csvsimple}
\usepackage{xcolor}
\usepackage{float}
\usepackage{stfloats}
\usepackage{graphicx}
\usepackage{graphbox}
\usepackage{chngcntr}
\usepackage{amsmath}
\usepackage{amssymb}
\usepackage{multirow}
\usepackage{verbatim}
\usepackage{multirow}
\usepackage[toc]{appendix}
\usepackage{soul}
\usepackage{txfonts}
\usepackage{natbib}
\usepackage[colorlinks,citecolor=blue,urlcolor=blue,bookmarks=false,hypertexnames=true]{hyperref} 
\bibpunct{(}{)}{;}{a}{}{,}
\makeatletter
\renewcommand*\aa@pageof{, page \thepage{} of \pageref*{LastPage}}
\makeatother

\defcitealias{Bourdin2017}{B17}

\begin{document}

   \title{Resolving high-z Galaxy Cluster properties through joint X-ray and Millimeter analysis: a case study of SPT-CLJ0615-5746}
   \subtitle{}
  
   \titlerunning{Multiwavelenght analysis of high-z galaxy cluster: SPT-CLJ0615-5746}
   %\subtitle{I. Overviewing the $\kappa$-mechanism}

   \author{C. Mastromarino
          \inst{1}\fnmsep\inst{2}
          \and
          F. Oppizzi\inst{1}\fnmsep\inst{3}
          \and
          F. De Luca\inst{1}\fnmsep\inst{2}
          \and
          H. Bourdin\inst{1}\fnmsep\inst{2}
          \and
          P. Mazzotta\inst{1}\fnmsep\inst{2}
          }
    
    \institute{Dipartimento di Fisica, Università di Roma ‘Tor Vergata’, Via della Ricerca Scientifica 1, I-00133 Roma, Italy.
    \and
    INFN, Sezione di Roma ‘Tor Vergata’, Via della Ricerca Scientifica, 1, 00133, Roma, Italy.
    \and
    INFN-Padova, Via Marzolo 8, 35131, Padova, Italy. \\
    \email{cmastromarino@roma2.infn.it}
    }

   \date{Received September 15, 1996; accepted March 16, 1997}

  \abstract
{We present a joint millimetric and X-ray analysis of hot gas properties in the distant galaxy cluster SPT-CLJ0615-5746 ($z = 0.972$).
Combining \textit{Chandra} observations with the South Pole Telescope (SPT) and \textit{Planck} data, we perform radial measurements of thermodynamical quantities up to a characteristic radius of $1.2\, R_{500}$.  We exploit the high angular resolution of \textit{Chandra} and SPT to map the innermost region of the cluster and the high sensitivity to the larger angular scales of \textit{Planck} to constrain the outskirts and improve the estimation of the cosmic microwave background and the galactic thermal dust emissions. Besides maximizing the accuracy of radial temperature measurements, our joint analysis allows us to test the consistency between X-ray and millimetric derivations of thermodynamic quantities via the introduction of a normalization parameter ($\eta_T$) between X-ray and millimetric temperature profiles. This approach reveals a substantial high value of the normalization parameter, $\eta_T=1.46^{+0.15}_{-0.22}$, suggesting that the gas halo is aspherical.\\
Assuming hot gas hydrostatic equilibrium within complementary angular sectors that intercept the major and minor elongation of the X-ray image, we infer a halo mass profile that results from an effective compensation of azimuthal variations of gas densities by variations of the $\eta_T$ parameter. Consistent with earlier integrated X-ray and millimetric measurements, we infer a cluster mass of $M^{\text{HE}}_{500} = 10.67^{+0.62}_{-0.50}\,\, 10^{14}\,M_{\odot}$.} 


   \maketitle
%
%-------------------------------------------------------------------

\section{Introduction}
Being the most massive gravitationally bound structure, galaxy clusters supposedly formed recently via collapse of matter haloes and accretion of substructures in their periphery.
As closed pieces of Universe, Galaxy clusters are laboratories that reveal us the nature of the matter content of the Universe, including dark matter and baryons, and the interplay between these component during the late time of the cosmic evolution. \\
Baryons in galaxy clusters prevalently take the form of a hot ionised atmosphere, detectable at the same time as a result of its inverse Compton scattering of CMB photons
(the so-called SZ effect \citet{sunyaev1972observations}) and in the X-ray band via its bremsstrahlung emission.  Due to the redshift independence of the SZ effect, millmetric surveys recently allowed us to sample the cluster population over a significant fraction of its evolution with cosmic times. In particular, the all-sky \textit{Planck} catalog of SZ sources \citep{ade2016planckb} contains 1203 confirmed clusters with a mean mass of $4.82\times 10^{14}M_{\odot}$ over a redshift range between 0.011 and 0.972. Instead, the first SPT receiver was employed to carry out the 2500-square degree SPT-SZ survey \citep{bleem2015galaxy}. The catalog features 677 cluster candidates, of which 516 have optical and/or infrared counterparts. The median mass of the catalog is $ 3.5 \times 10^{14}M_{\odot}h^{-1}_{70}$ with a median redshift of $z=0.55$, while the highest-redshift systems reach $z>1.4$. The second receiver has been used to conduct the 500-square degree SPTpol survey \citep{bleem2023galaxy}, yielding a new catalog containing 689 detections. Out of these, 544 are confirmed clusters with identified counterparts in optical and infrared data. The confirmed sample has a median mass of $2.5 \times 10^{14} M_{\odot}h^{-1}_{70}$ and a median redshift of $z=0.7$. The highest-redshift clusters extend up to $z\sim1.6$, with 21\% of the samples at $z>1$. X-ray observations will ideally complement millimetric observations in understanding the hot gas thermodynamics in these newly detected clusters. This is particularly significant since the SZ effect and thermal bremsstrahlung radiation are predominantly sensitive to the hot gas pressure and density, respectively, thus a combination of the two measurements would allow a direct estimation of the hydrostatic mass and the other thermodynamic profiles \citep{ruppin2021stability, adam2023xxl}. However, cross-calibration issues of the instruments, the morphology of the systems, or the cosmological model can arise discrepancies between the two independent observations of the hot gas. Such systematics can be accounted through the introduction of a normalization parameter, $\eta_T$,  between X-ray and millimetric temperature profiles, enabling the effective combination of the two datasets. In the ideal scenario, we would expect $\eta_T=1$, but in more realistic cases $\eta_T$ is expected to deviate from one due to the various factors mentioned earlier \citep{bourdin2017pressure,ghirardini2019universal,kozmanyan2019deriving}.

To investigate the ICM thermodynamics in high redshift clusters, we present a joint millimetre and X-ray analysis of hot gas properties of SPT-CLJ0615-4756, the most distant galaxy cluster in the Planck catalogue of SZ sources (\textit{Planck} name: PSZ2 G266.54-27.31). This cluster is also included in the SPT-SZ cluster catalogue with a mass estimates of $M_{500}= 10.16^{+0.98}_{-1.20}\,\, 10^{14} M_{\odot}$, making it one of the most massive clusters in the survey. Studies of this cluster have been conducted in various wavelengths. Optical research \citep{schrabback2018cluster, connor2019assembling} and X-ray analysis \citep{bartalucci2017resolving,bartalucci2018resolving} have both revealed an elongated shape and a high core-excised X-ray temperature. These results are in agreement with the large $\eta_T$ value reported by \citet{bourdin2017pressure} ($\eta_T \sim 1.7$), which suggests a high level of asphericity. Recently, \citet{jimenez2023dissecting} used the intracluster light (ICL) fraction as an indicator of the dynamical stage of the cluster and found that the ICL fraction distribution of SPT-CLJ0615-4756 is more indicative of an active system than a relaxed cluster.

In this work, we present an updated version of the pipeline developed in \citet{bourdin2017pressure} and \citet{oppizzi2023chex} to analyse the properties of such high redshift cluster. We take advantage of the multi-wavelength data from \textit{Chandra} X-ray Observatory, \textit{Planck}, and the South Pole Telescope (SPT) to uncover the structure of the ICM out to distances beyond $R_{500}$. \\
We present the three data sets in Section \ref{dataset} and their processing in Section \ref{dataprocessing}. We discuss our fitting methods in Section \ref{fittingprocedure}, with the parametric modeling and  analysis of the ICM properties in Section \ref{Result}. Our summary and conclusions are presented in Section \ref{Summaryandconclusion}. In this paper, we assume a flat $\Lambda$CDM cosmology with $\Omega_m = 0.3$, $\Omega_{\Lambda}=0.7$, and $H_0 = 70 \,km\,s^{-1}\,Mpc^{-1}$. We define $R_{500}$ and $M_{500}$ in terms of the critical density $\rho_c(z)$ at cluster redshift $z$ as $M_{500} = \frac{4\pi}{3} \,500\rho_c(z) R_{500}^3$.

\section{Data-sets}
\label{dataset}
SPT-CLJ0615-5746 (\textit{Planck} name: PLCKG266.6-27.3) is the highest redshift cluster detected by \textit{Planck} via SZ effects \citep{ade2016planckb}. This system also belongs to the SPT-SZ cluster sample \citep{bleem2015galaxy} and has additionally been observed with \textit{Chandra} with an exposure time $\sim 240ks$ (proposal 13800663, PI P. Mazzotta). The latter observations have been taken using the Advanced CCD imaging spectrometer (ACIS, \citet{garmire2003advanced}) and are part of the public data available in the \textit{ Chandra} data archive\footnote{\url{https://cda.harvard.edu/chaser/}}.\\
\textit{Planck} data come from the second public data release\footnote{\url{https://pla.esac.esa.int/}} (PR2) of \textit{ Planck} High-Frequency Instrument (HFI). The dataset consisted of six full-sky maps in HEALPIX format with an $N_{\text{side}}$ value of 2048, meaning a pixel angular size of 1.72 arcmin. These maps were obtained from the full 30-month mission and had nominal frequencies of 100, 143, 217, 353, 545, and 857 GHz, with resolution 9.66, 7.22, 4.90, 4.92, 4.67, 4.22 arcmin FWHM Gaussian, respectively.\\
In addition, we use the public SPT data\footnote{\url{https://pole.uchicago.edu/public/data/chown18/index.html}} from 
the 2500-square-degree SPT–SZ survey from 20h to 7h in right ascension and from $-65^{\circ}$ to $-40^{\circ}$ in declination and consist of three maps with frequency bands centered at $95,\,150\,\text{and}\,220\, GHz$ with $1.7,\,1.2\,\text{and}\, 1.0$ arcmin resolution, respectively. These public data are convolved with a common Gaussian beam with $1.75\, \rm arcmin\rm$ FWHM and released in the HEALPIX format with resolution parameter $N_{\text{side}}= 8192$ corresponding to a pixel angular size of 0.43 arcmin. The main properties of the cluster and our X-ray data set information are listed in Tab. \ref{tab:properties}.
\begin{table}
	\centering
	\caption{Main properties of SPT-CLJ0615-5746. The cluster coordinates correspond to the X-ray peaks. $M^{\text{SPT}}_{500}$, redshift and the signal to noise $S/N$ come from the SPT catalog \citep{bleem2015galaxy}, we use these values to derive $R^{\text{SPT}}_{500}$. The published SPT masses are the estimated 'true' mass from the SZ signal, while the \textit{Planck} masses are derived from the $Y_{\text{SZ}}-M_{500}$ relation and they are not corrected for hydrostatic bias.}
   \begingroup

    \setlength{\tabcolsep}{12pt} % Default value: 6pt
    \renewcommand{\arraystretch}{1.6} % Default value: 1
	\begin{tabular}{cc}
		\hline
            \hline
		\multicolumn{2}{c}{SPT-CLJ0615-5746} \\
            \hline
            \hline

		  X-ray peak (R.A., Dec.) &93.9663 \, -57.7797  \\
            \hline

		z & 0.972 \\ 
            \hline
            $M^{\text{SPT}}_{500}\,[10^{14}M_{\odot}]$ & $10.16^{+0.98}_{-1.20}$ \\ 
            \hline
            $R^{\text{SPT}}_{500}\,[arcmin]$ & $2.22^{+0.07}_{-0.09}$ \\ 
            \hline
            $R^{\text{SPT}}_{500}\,[\rm kpc\rm]$ & $1059.34^{+34.06}_{-41.71}$ \\ 
            \hline
            $S/N$ &  26.4 \\         
		\hline
            \hline
	\end{tabular}
 \endgroup
	\label{tab:properties}\centering
\end{table}

\section{Data processing}
\label{dataprocessing}
\subsection{Chandra observations}
\begin{figure}
    \centering
    \includegraphics[width=0.9\linewidth]{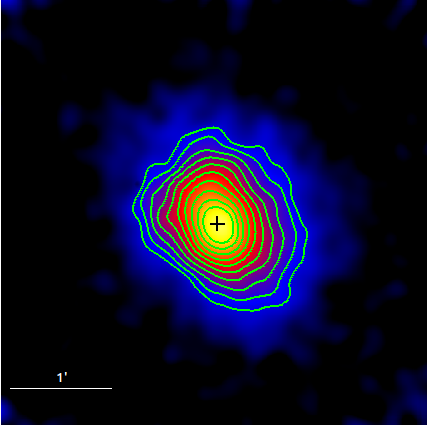}
    	\caption{Wavelet-denoised image of the \textit{Chandra} X-ray emission of SPT-CLJ0615-5746. The map shows the backgroud-subtracted surface brightness, computed in the energy range $[0.5,2.5]\,\rm keV$ and covers an area of $2R^{\text{SPT}}_{500}\times2R^{\text{SPT}}_{500}$ from the position of the X-ray peak (black cross). The surface brightness isocontour levels (green lines) are logarithmically equispaced by a factor $\log \sqrt{2}$. }
	\label{fig:wavletmap}
\end{figure}
We analyse the X-ray data following the scheme used in \citet{bourdin2017pressure}.
The CDFS from the \textit{Chandra} Data Archive are first pre-processed to generate calibrated event files for the data reduction with the CIAO tools, version 4.13 and calibration files version 4.9.6. To prevent any contamination from flares or high-background periods, we follow \citet{bourdin2008temperature,bourdin2013shock} and remove all the events that deviate more than $3\sigma$ from the light curve profile. We identify and mask all point sources that can contaminate our signals using SExtractor \citep{bertin1996sextractor}.\\
The X-ray background of the \textit{Chandra} telescope is characterized by instrumental (particle background) and astrophysical components.
To mitigate potential systematic errors arising from residual spatial fluctuations of the particle background model across the entire Field of View, we fit the normalization of the particle background model in a sky region that surrounds the cluster itself. This involves considering all pixels between $0.35\,R^{\text{SPT}}_{500}$ and $0.8\,R^{\text{SPT}}_{500}$ from the X-ray peak, using the $[9.5-10.6],\rm keV$ band. To ensure that contamination from the cluster is not considered, we extract the spectrum within the specified radial range, calculating a Signal-to-Noise Ratio (SNR) of approximately 4\% in our energy band. This low SNR is expected, as at these energies, the ACIS-I effective area decreases by a factor of approximately 100, resulting in cluster counts comparable to the level of systematic in the background estimation. This approach allows us to accurately estimate the background surrounding the observed cluster while simultaneously minimizing the potential risk of contamination from the cluster itself. The astrophysical component originates from the foreground emission of our Galaxy and the cosmic X-ray background (CXB) as a result of the unresolved point sources. To have high photon statistics, we fit the normalization of these components considering all the pixels that lie more than $1.5\,R^{\text{SPT}}_{500}$ from the X-ray peak following the spectral and spatial features described in \citet{bourdin2013shock}: the CXB is modeled with an absorbed power law of index $\gamma =1.42$, while the emission associated with our Galaxy can be modeled by the sum of two absorbed thermal components accounting for the Galactic transabsorption emission ($kT1 =0.099\, \rm keV\rm$ and $kT2=0.248\, \rm keV\rm$).\\
Figure \ref{fig:wavletmap} shows the zoomed wavelet map after the cleaning procedure. The image is produced using a denoising algorithm based on a wavelet reconstruction of the maps \citep{starck2009source, bourdin2013shock}. The map is background subtracted and covers a region of $2R^{\text{SPT}}_{500}\times2R^{\text{SPT}}_{500}$ with the X-ray surface brightness extracted in the soft X-ray band (0.5, 2.5\,keV). The black cross represents the X-ray peak, identified as the coordinates of the maximum of a sparse wavelet-denoised \citep{starck2002deconvolution,starck2009source} surface brightness map. We consider this position as the cluster center to compute all the radial profiles for both X-ray and SZ analysis.
\subsection{Millimetric observations}
\begin{figure*}
    \centering
    \includegraphics[width=0.33\linewidth]{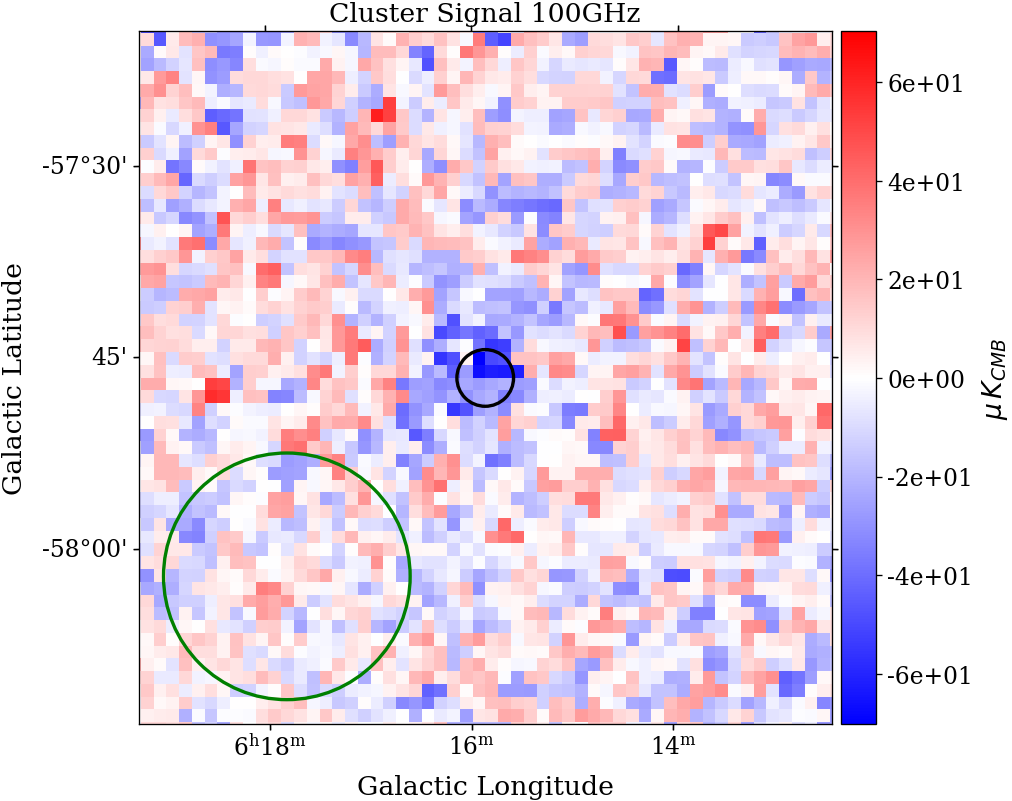}
    \includegraphics[width=0.33\linewidth]{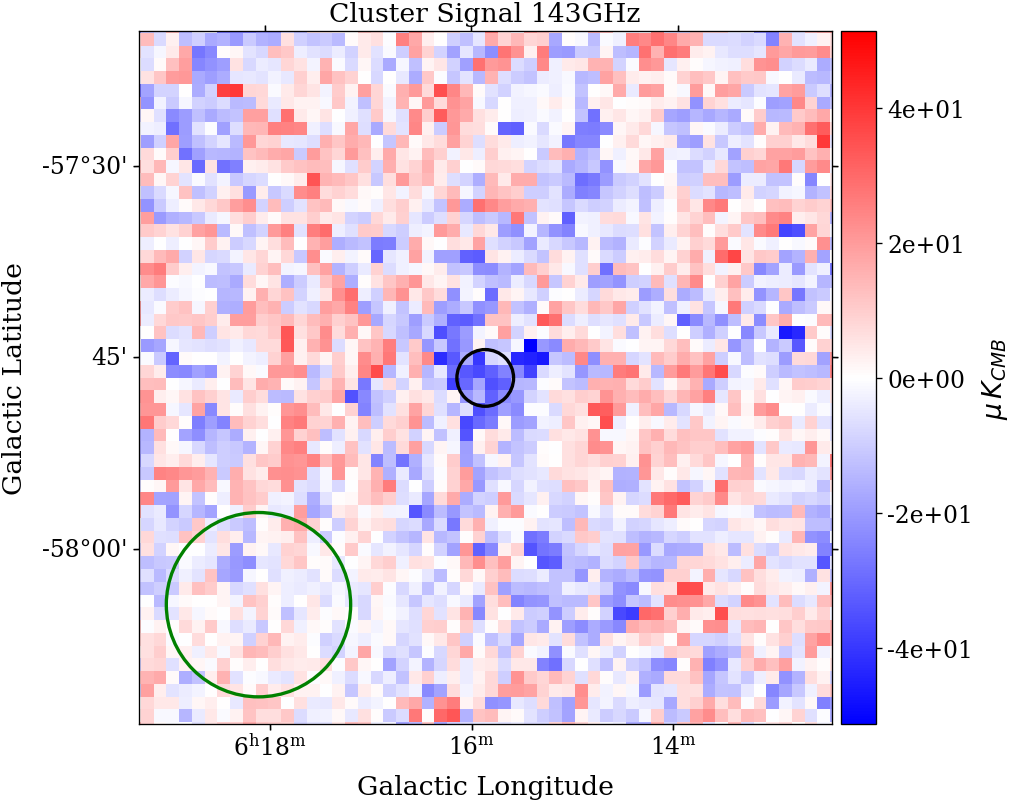}
    \includegraphics[width=0.33\linewidth]{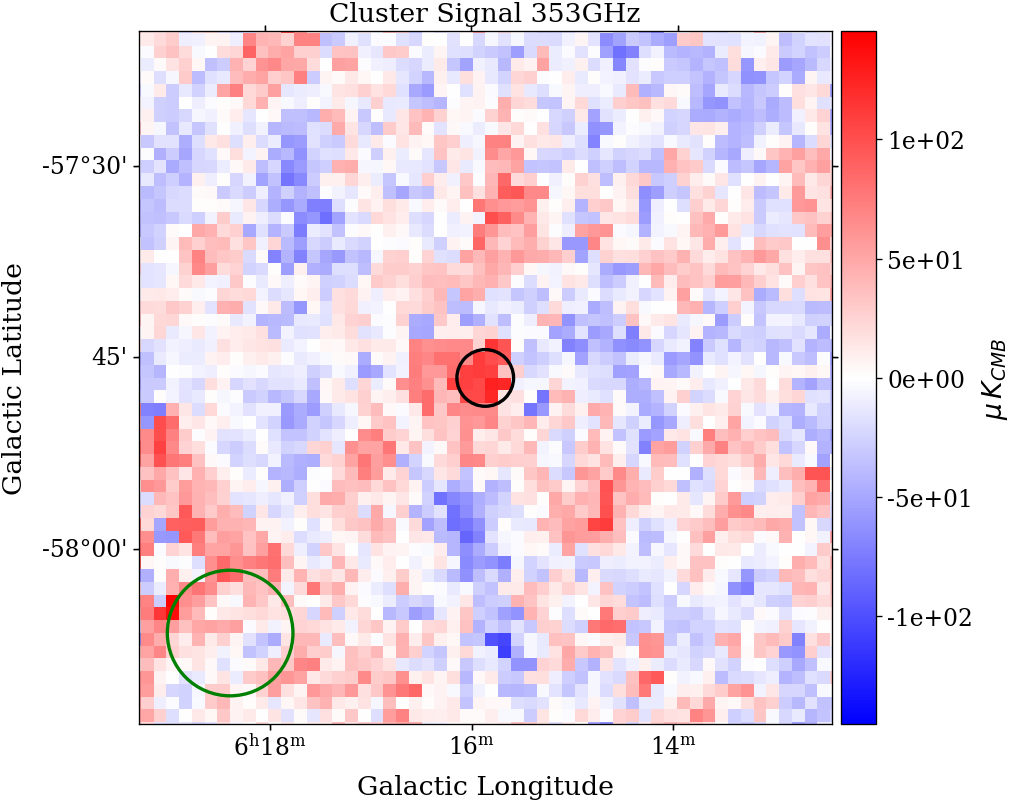}
    	\caption{Planck cleaned maps of SPT-CLJ0615-5746 of size $12\,R^{\text{SPT}}_{500}\times12\,R^{\text{SPT}}_{500}$ from the $100\,GHz$ (left), $143\,GHz$ (center) and $353\,GHz$ (right) channels. The black circles mark $R^{\text{SPT}}_{500}$ and the green circles are the beam FWHM for each channel. The pixel size is $1\,arcmin$. }
	\label{fig:Planckmaps}
\end{figure*}
\begin{figure*}
    \centering
    \includegraphics[width=0.4\linewidth]{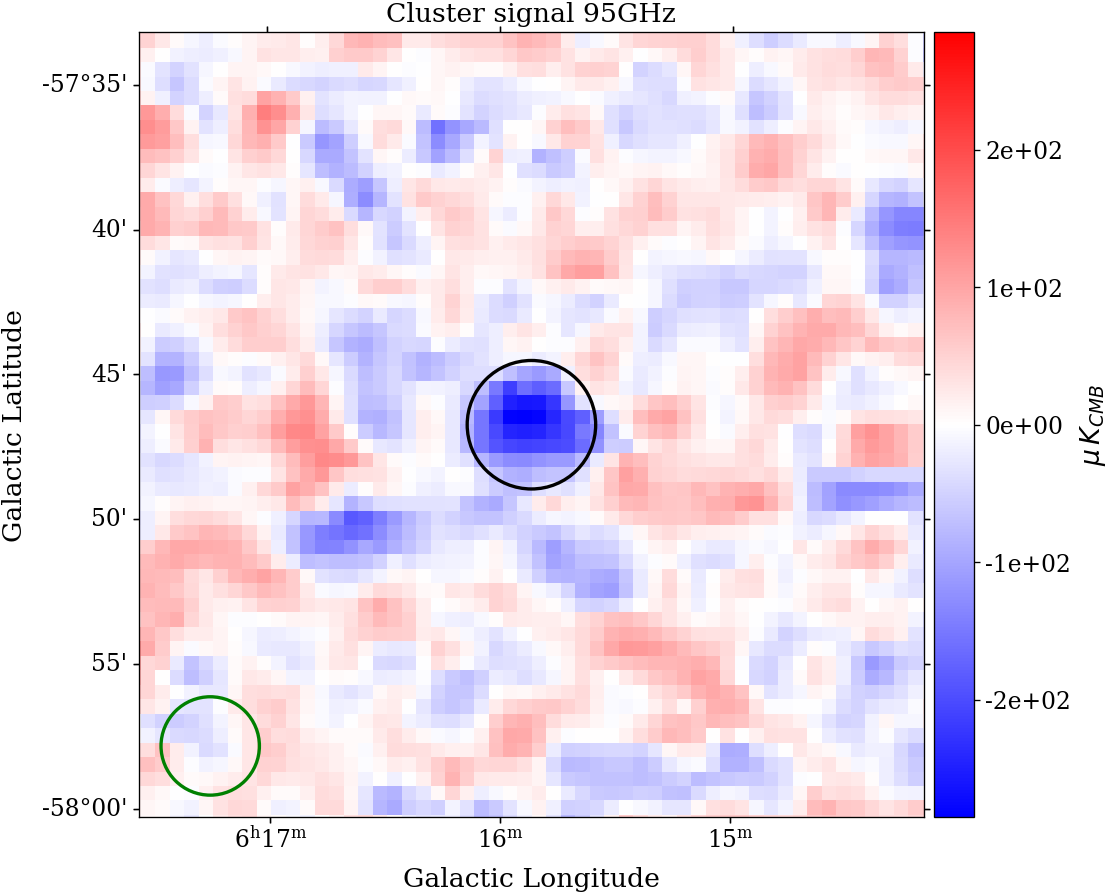}
    \includegraphics[width=0.4\linewidth]{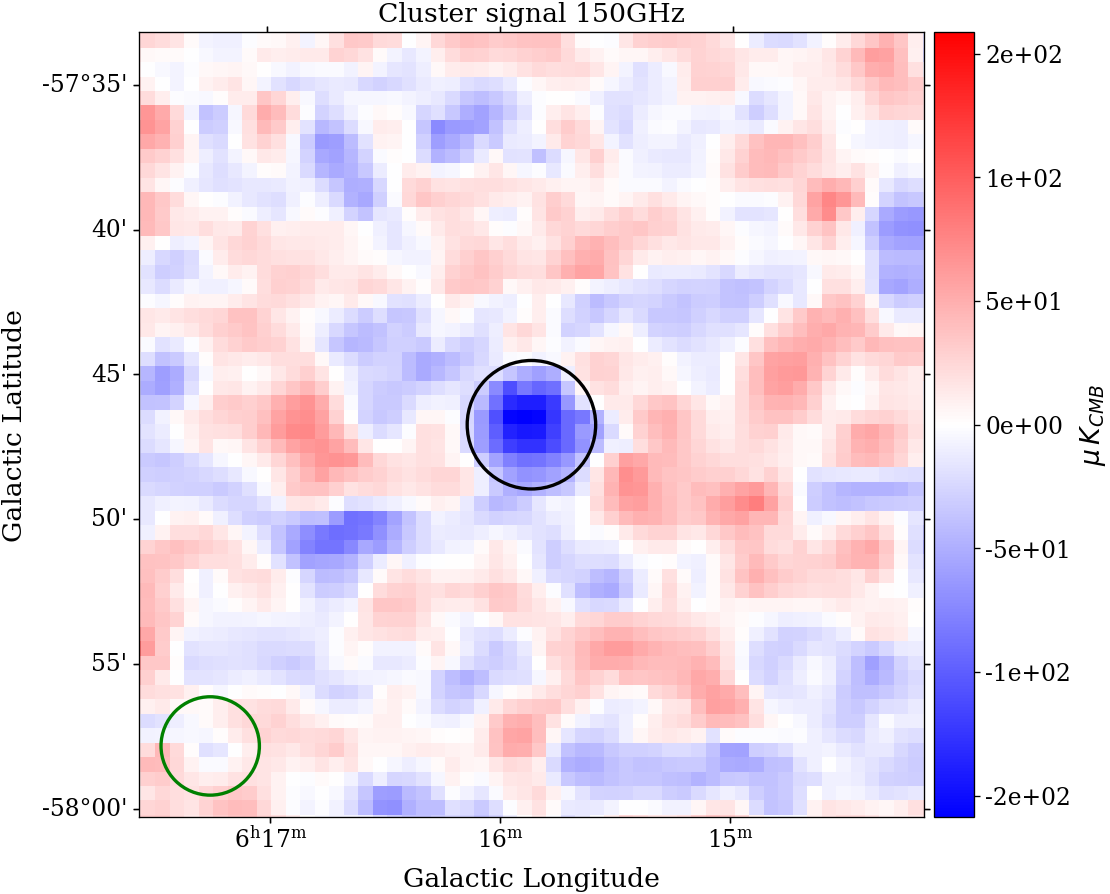}
    	\caption{SPT cleaned maps of SPT-CLJ0615-5746 of size $6\,R^{\text{SPT}}_{500}\times6\,R^{\text{SPT}}_{500}$ from the $95\,GHz$ (left) and $150\,GHz$ (right) channels. The black circles mark $R^{\text{SPT}}_{500}$ and the green circles are the beam FWHM for each channel. The pixel size is $0.5\,arcmin$. }
	\label{fig:sptmaps}
\end{figure*}
To analyse the millimetric data and extract the pressure profile of the cluster SPT-CLJ0615-5746, we use an updated version of the pipeline described in \citet{oppizzi2023chex}. This new pipeline essentially consists of two steps using both SPT and \textit{Planck} maps: a component separation to remove the background (CMB background) and the foreground (Galactic contaminants) in the SPT and \textit{Planck} observations separately and then a combination of the two datasets to derive the pressure profile.\\
\subsubsection{Planck observations}
The galaxy cluster signal has been extracted using \textit{Planck} HFI data from the full 30-months mission. The all-sky HFI frequency maps have been reprojected on smaller tiles of $512\times512$ pixels into the tangential plane, with a resolution of $1\,\rm arcmin/pixel\rm$ (corresponding to
$\sim 8.5^{\circ}\times8.5^{\circ}$) around the cluster X-ray peak. As a first consideration, we remove all point sources from the HFI maps, masking all the ones detected in the PCCS2 and PCCS2E catalogs \citep{ade2016planck}. These maps incorporate the SZ and dust cluster signal combined with Galactic foreground and CMB anisotropies. Additionally, other contributions come from cosmic infrared and SZ backgrounds, along with instrumental offsets. To isolate the millimetric cluster signal from the other components, we use the technique developed in \citet{bourdin2017pressure}. Here we summarize the pipeline: 
\begin{enumerate}
    \item We reduce the angular resolution of HFI maps in the range $217-857\,\rm GHz\rm$, to a common value of $5\,\rm arcmin\rm$ and then, to remove the large scale fluctuations, we convolve each frequency map with a third-order B-spline kernel \citep{curry1965polya}. This procedure yields a smoothed image that we subtract from the raw image to get the maps filtered from large-scale anisotropies.
    \item To create a spatial template for the Galactic thermal dust anisotropies (GTD), we denoise the $857\,\rm GHz\rm$ map using an isotropic undecimated wavelet transform \citep{starck2007undecimated} and applying a thresholding ($3\sigma$) on the coefficients.
    Then we model the SED of these anisotropies as a two temperature ($T_{\text{cold}}$ and $T_{\text{hot}}$) gray body \citep{meisner2014modeling}. Defining $\beta_{\text{d,cold}}$ and $\beta_{\text{d,hot}}$ as spectral indexes, $f_{\text{cold}}$ as the cold component fraction and $q_1$ and $q_2$ the ratios of far infrared emission to optical absorption, the SED is as follows:
    \begin{multline}
        s_{GTD}(\nu)=\Bigg[\frac{q_1f_{\text{cold}}}{q_2}\Bigg(\frac{\nu}{\nu_0}\Bigg)^{\beta_{\text{d,cold}}}B_{\nu}(T_{\text{cold}})+\\
        +(1-f_{\text{cold}})\Bigg(\frac{\nu}{\nu_0}\Bigg)^{\beta_{\text{d,hot}}}B_{\nu}(T_{\text{hot}})\Bigg]\,,
    \end{multline}
    where $B_{\nu}$ is the Planck function. Here, we leave free the overall amplitude, $f_{\text{cold}}$ and $\beta_{\text{d,cold}}$, while the other parameters are fixed from their average all-sky values. Moreover, $T_{\text{hot}}$ and $T_{\text{cold}}=f(T_{\text{hot}},q_1/q_2,\beta_{\text{d,cold}},\beta_{\text{d,hot}})$ are mapped a priori from a joint fit to \textit{Planck} and IRAS all-sky maps \citep{finkbeiner1999extrapolation, meisner2014modeling}.
    \item To characterize the spatial CMB template, we subtract the GTD template from the denoised $217\,\rm GHz\rm$ map (as for the $857\,\rm GHz\rm$ map). Both the GTD and CMB templates are then rescaled to all HFI frequencies and jointly fitted to the data extracted in the cluster-centric radii range $[7,12]R^{\text{SPT}}_{500}$, where no cluster signal is expected.
    \item Finally, we subtract these templates from the HFI maps, getting the cluster thermal SZ signals with a residual contribution from the cluster thermal dust emissivity (CTD).
\end{enumerate}
Figure \ref{fig:Planckmaps} shows the clean frequency maps that we use in the fit in Section \ref{millimetricfit}.

\subsubsection{SPT observations}
Similary to the \textit{Planck} maps, we use the public SPT maps, reprojected on smaller tiles of $1024\times1024$ pixels into the tangential plane, with a resolution of $0.5\,\rm arcmin/pixel\rm$ (corresponding to $\sim 8.6^{\circ}\times8.6^{\circ}$) around the X-ray peak of the cluster. We mask all point sources detected in the SPT maps, following the procedure detailed in \citet{chown2018maps}.\\
Due to the differences in frequencies and spatial scales covered, the SPT data are not compatible with the component separation technique explained in the previous section, which was specifically designed for the \textit{Planck} data. Specifically, the SPT lacks high-frequency channels necessary for deriving the dust templates essential for the multi-component fit described earlier. Additionally, the noise level in the SPT's $220\,GHz$ channel is too high to accurately recover the CMB templates with the required precision. Without the templates to fit, \citet{oppizzi2023chex} developed a new method tailored to the SPT data structure, including information from \textit{Planck} to improve the reconstruction on larger spatial scales. Here we summarize the pipeline:
\begin{enumerate}
    \item Similarly to what has been done in previous work \citep{delabrouille2009full,remazeilles2011foreground,hurier2013milca,oppizzi2020needlet}, to recover the CMB signal, we start from a linear combination (LC) of the three SPT frequency maps and the $217\,\rm GHz\rm$ map from \textit{Planck}. Since the SPT window function changes with the channels, to make a LC of them, we first equalize the maps, including the \textit{Planck} one, to the $150\,\rm GHz\rm$ spatial response.
    \item To combine maps from the two instruments with different resolution, we split each SPT map in a low-pass filtered map and in a high-pass filtered map.
    \item The LC weights are then calculated on a subregion of $1^{\circ}\times 1^{\circ}$ around the X-ray peak, separately on the low- and high-passed maps (with the addition of the \textit{Planck} map). In both cases, these are calculated as a double optimization: to minimize the variance with respect to a signal constant in frequency (CMB) and simultaneously to null the non-relativistic SZ component. These conditions lead to the following:
    \begin{equation}
        \textbf{w}=\frac{eC^{-1}}{A^TC^{-1}A}\,,
    \end{equation}
    where \textbf{w} are the weights assigned to each frequency, $C$ is the data covariance matrix between the $N_{chan}$ channels, $e$ is a vector of lengths $N_{chan}$ and $A$ is the matrix accounting for the emission of the CMB and SZ signal at the various frequencies.
    \item We recombine the CMB estimations of the large and small scales into a single map and rescale to all SPT frequency to obtain the final templates.
    \item Finally, we subtract the templates from the SPT maps, obtaining the clean maps with the cluster signal.
\end{enumerate}
Figure \ref{fig:sptmaps} shows the clean frequency maps that we use in the fit in Section \ref{millimetricfit}.
\section{Hot gas thermodynamics and profile fitting}
\label{fittingprocedure}
With the case of SPT-CLJ0615-5746, we take advantage of the opportunity to validate the SZ profile fitting method at high redshifts and test if there is consistency between X-ray and millimetric data when incorporating \textit{Chandra} data. In our analysis, we employ and compare two different fitting methods. The first approach involves the separate examination of the X-ray and millimeter observable. The second method incorporates a multi-wavelength joint fit that combine data from three different instruments. This approach is designed to harness the strengths of each instrument and address potential systematic effects that may arise from the integration of \textit{Chandra} data with SPT-\textit{Planck} data, thereby providing a more robust and comprehensive analysis.
\subsection{X-ray Fit}
\label{X-ray-Fit}
\begin{figure}
    \centering
    \includegraphics[width=1\linewidth]{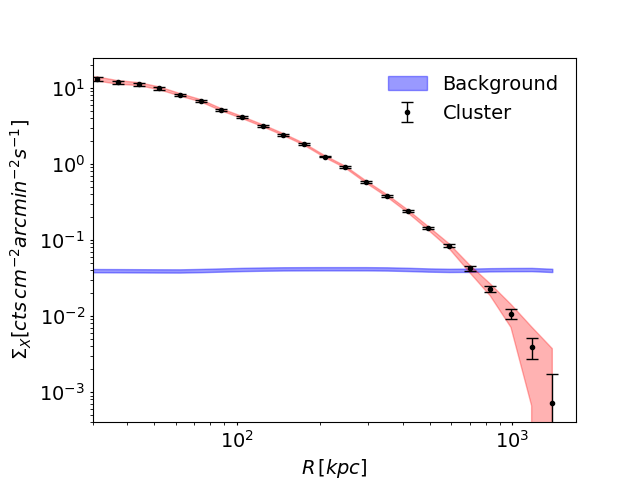}
    \caption{Background-subrtacted surface brightness profile of the Chandra data (black points) measured on SPT-CLJ0615-5746. The black bars represent the Poissonian statistical error associated with the measurement in each bin, while the shaded red regions represent the sum of the statistical and the 5\% systematic error associated with the estimation of the background X-ray emission (shaded blue region).}
    \label{fig:Brightness}
\end{figure}
\begin{figure*}
    \centering
    \includegraphics[width=0.47\linewidth]{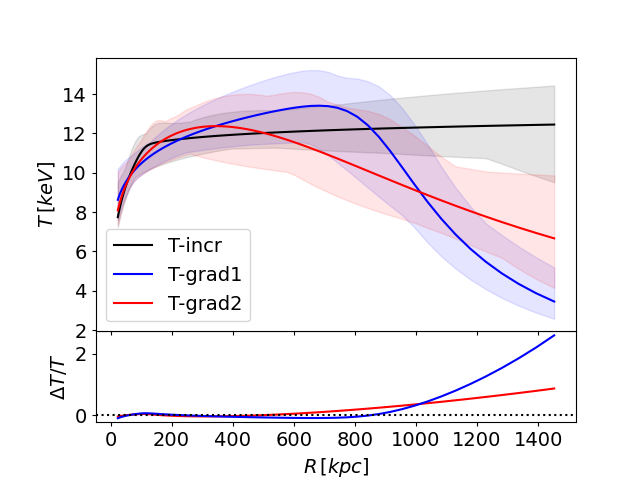}
    \includegraphics[width=0.47\linewidth]{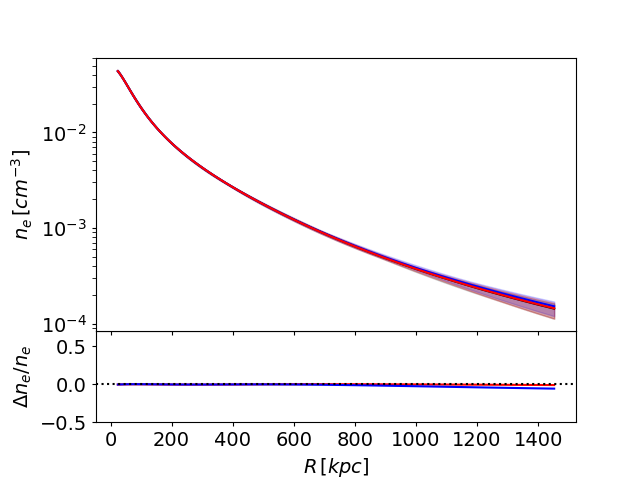}
    \caption{Left: 3D temperature profiles for three different scenario. The black line represents an increasing temperature in the outskirts, while the blue and red lines represent two decreasing temperatures with different gradients. Right: Respective density profile for each scenario. The shaded regions correspond to the 68\% credible intervals. The bottom panels represent the relative deviation with respect to the increasing temperature (black profiles).}
    \label{fig:diff_temperature}
\end{figure*}

The cluster emission from the hot plasma present in the ICM is modeled by combining the bremmsstrahlung continuum and the metal emission lines with the APEC spectral library and the absorption due to the Galactic medium. For the latter, we consider the photoelectric cross-sections from \citet{verner1995atomic}, the abundance table of \citet{asplund2009chemical}. We consider the total hydrogen column density defined as the sum of the atomic hydrogen and the molecular contributions. We estimate this value directly from the X-ray spectrum of the cluster, considering a circular annulus with radii $[0.15-0.8]\,R^{SPT}_{500}$ thus excluding the central regions. We perform the fit in the energy band $[0.5-10.0]\,\rm kev\rm$ and left free to vary the total hydrogen column density, the temperature, the metallicity and the normalization of the spectrum. We find $N_H = 4.56_{-0.68}^{+0.55}  \times 10^{20}\, cm^{-2}$ that is consistent with the value reported in the public data from the LAB survey \citep{kalberla2005leiden} of $N_H = 4.32 \times 10^{20}\, cm^{-2}$.
Using the X-ray peak, we extract the radial profiles of the background-subtracted X-ray observable: the surface brightness ($\Sigma_X$) estimated in the soft X-ray band ($[0.5,2.5]\,\rm keV\rm$) and the spectroscopic temperature ($T_X$) considering the energy band $[0.5,10]\,\rm keV \rm$. The former is defined as:
\begin{equation}
     \Sigma_X(r) =\frac{1}{4\pi(1+z)^3}\int \frac{n_e^2(r)}{x}\Lambda(T,Z)dl \,,
    \label{eq:Xemission}
\end{equation}
where $z$ is the redshift of the cluster and $\Lambda(T,Z)$ is the cooling function of the cluster emission dependent on the temperature T and the element abundances Z. Surface brightness is expressed in detector units (photon counts $\rm cm^{-2} arcmin^{-2} s^{-1}\rm$).
For $T_X$ we assume the temperature weighting scheme proposed in \citet{mazzotta2004comparing}, to mimic the spectroscopic response of a single temperature:
\begin{equation}
    T_X(r) = \frac{\int WT(r)\,dl}{\int W\, dl}\,,
    \label{mazzottaequation}
\end{equation}
with $W = n_e^2/T^{3/4}$.\\
We then parameterize the emission measure $[n_pn_e](r)$ using analytical forms first proposed in \citet{vikhlinin2006chandra}: 
\begin{multline}
    n_pn_e(r) = \frac{n_o^2(r/r_c)^{-\alpha}}{[1+(r/r_s)^{\gamma}]^{\epsilon/\gamma}[1+(r/r_c)^2]^{3\beta_1 -\alpha/2}} +\\ 
    + \frac{n_{0,2}^2}{[1+(r/r_{c,2})^2]^{3\beta_2}}\,,
    \label{density}
\end{multline}
where $\alpha$ is the cusp index accounting the higher densities in the cluster centre, $n_0$ and $n_{0,2}$ are the normalisation of two $\beta$-model profiles, with slopes $\beta_1$ and $\beta_2$ and characteristic radii $r_c$ and $r_{c,2}$, respectively. At large radii, $\epsilon$ represents the asymptotic change in slope near a given radius $r_s$, with $\gamma$ ruling the width of the transition region.
For the temperature profile $T(r)$, we use the functional form suggested in \citet{vikhlinin2006chandra}:
\begin{equation}
    \label{3Dtemperature}
    T(r) = T_0 \frac{x+ T_{min}/T_0}{x+1}\frac{(r/r_t)^{-a}}{[1+(r/r_t)^b]^{c/b}}\,,
\end{equation}
where $T_0$ is the overall normalisation. The first ratio describes the decrease in temperature in the center as a function of the typical cooling scale $x=(r/r_{coo})^{a_{cool}}$ to a minimum temperature $T_{min}$. The second is a broken power law, with characteristic scale $r_t$ and transition region describing the intermediate and outskirt profile of the cluster, with slope $a$, $b$ and $c$.\\
These templates are then integrated along the line of sight and fitted to the two observable profiles ($\Sigma_X$ and $T_X$). All fits are carried out with least-squares minimization following the Levenberg–Marquardt algorithm. For uncertainties envelopes of the best-fit profiles, we perform 500 parametric bootstrap realizations of the observed profiles, estimated considering 8 radial bins inside $1.4\, R^{\text{SPT}}_{500}$ for the spectroscopic temperature, and 25 logarithmic radial bins for the surface brightness inside $1.4\, R^{\text{SPT}}_{500}$. 
We show the surface brightness profile of our system in Figure \ref{fig:Brightness}. The black dots represent the background subtracted data with their respective Poissonian statistical error associated with the measurement in each bin. The horizontal blue bar represents the background model with its $5\%$ systematic error (details of the systematic errors associated with \textit{Chandra} observations are in \citet{bartalucci2014chandra}). The shaded red regions represent the sum of the statistical and systematic errors associated with the estimation of the background X-ray emission. In the last two bins we found that the emission measurements are more than 10 times below the background with the corresponding high dispersion, making the result for $R> 1200\,\rm kpc\rm \sim 1.2\,R^{\text{SPT}}_{500}$ less reliable. Consequently, we do not employ the results beyond these radii in direct comparisons.\\
The decision to employ this radial range is motivated by our goal of derive the thermodynamic and hydrostatic mass profiles extending to the largest possible radii. However, at such high redshifts, obtaining a spectroscopic temperature measurement in the outer regions of the cluster presents a significant challenge. Specifically, we find that in the last two bins corresponding to $[0.8,1.0]R^{\text{SPT}}_{500}$ and $[1.0,1.4] R^{\text{SPT}}_{500}$, the signal-to-noise ratio is merely $12.4\%$ and $5.4\%$, respectively. This implies that we lack sufficient photon counts to reliably infer the temperature in these regions. Consequently, we are unable to use the X-ray spectroscopic temperature data in the outermost portions of the cluster to extract our profiles.\\
In this context, our primary goal is to constrain the X-ray density profile up to a high radial distance. To convert the emission measure profiles to gas density, we need the values of a factor that depend on temperature and element abundances ($\Lambda(T,Z)$). However, if the temperature is not well constrained, it may lead to inaccuracies in the brightness profile and consequently affect the accuracy of the density profile. To assess the sensitivity of the density profile to temperature variations at high radii, we conduct an analysis to evaluate how different temperature values in these outer regions can impact the final density profile.We deliberately manipulate the spectroscopic temperature values in the last two bins, creating three distinct scenarios: one with increasing temperature after $0.8R^{\text{SPT}}_{500}$ (represented by the black line in the left panel of Figure \ref{fig:diff_temperature}), and two with decreasing temperatures with different gradients (shown as the blue and red lines in the left panel Figure \ref{fig:diff_temperature}). We then perform the fit on the surface brightness profile using these different temperature profiles.\\
The results of this test, as shown in the right panel of Figure \ref{fig:diff_temperature}, indicate that even extreme temperature differences in the outer regions have a minimal impact, causing less than a 5\% variation in the density profile within the same regions. This suggests that we can effectively constrain the density profile up to $R>R^{\text{SPT}}_{500}$ without requiring precise knowledge of the exact temperature profile.\\
Finally, adopting a metal abundance constant normalization ($Z=0.3$) for the cluster, the primordial helium abundance $Y_P=0.243$ from \citet{aghanim2020planck} results, and a particle mean weight of $\mu = 0.596$ allows us to infer an average ionization fraction $n_p/n_e = 0.852$. We can use this value to infer the electronic density distribution $n_e(r)$, from the parametric form of $[n_pn_e](r)$ (Eqn. \ref{density}). The upper left panels of  Figure \ref{fig:temp_and_entropy} shows the best-fit deprojected 3D density profile with its corresponding 68\% uncertainty envelope.

\subsection{Millimetric Fit}\label{millimetricfit}
To probe the gas pressure structure in both the inner and outer cluster regions, we combine the SPT and \textit{Planck} data in a joint fit. We first assume that the pressure structure follows the spherically symmetric profile proposed by \citet{nagai2007effects}:
\begin{equation}
    P(r)=P_0\times \frac{P_{500}}{x^{\gamma}(1+x^{\alpha})^{(\beta-\gamma)/\alpha}}\,,
    \label{pressure}
\end{equation}
where $x=c_{500}r/R_{500}$, while $\gamma,\alpha$ and $\beta$ are the slopes at $ r \ll R_{500}/c_{500},\,r\sim R_{500}/c_{500}\,\text{and}\, r \gg R_{500}/c_{500}$ respectively. To fit the millimetric data, $P(r)$ is used to build the thermal SZ signal expected for each channel in both \textit{Planck} and SPT. We project the 3D profile along the line of sight, and then convert in SZ brightness:
\begin{equation}
    I_{\text{SZ,c}}=s_{\text{SZ,c}}\frac{\sigma_T}{m_ec^2}\int P(r)dl \,.
    \label{ISZ}
\end{equation}
The $s_{\text{SZ,c}}$ coefficients that represent the frequency scaling, are derived from the non-relativistic Kompaneets equation:
\begin{equation}
    s_{\text{SZ,c}}=\int\,d\nu\,R_c(\nu)x(\nu)\Bigg[\frac{e^{x_{\nu}}+1}{e^{x_{\nu}}-1}-4\Bigg]\,,
\end{equation}
where $x(\nu)= h\nu/kT$, while $R_c(\nu)$ define the channel spectral response that for \textit{Planck} is given by the HFI model \citep{adam2016planck}. For SPT the spectral response is given in \citet{chown2018maps}.\\
For \textit{Planck} observations we also add a correction term to take into account the CTD component, that has been observed in \textit{Planck} data \citep{,collaborartion2016planck,adam2016planck}. \\
The complete model is then applied to fit the frequency maps from both SPT and \textit{Planck} to derive the amplitudes of the templates and contaminant components on a channel-by-channel basis while marginalizing over the cluster signal. Clean cluster maps, obtained by differencing the background and foreground templates, are used for this fitting process.

For each cleaned map we calculate the radial SZ profiles centered on the X-ray peak. Profile values are calculated by averaging over the pixels whose centers fall within the annulus defined by the bin edges. We employ 11 bins up to $3\,R^{\text{SPT}}_{500}$ for SPT and 8 bins up to $5\,R^{\text{SPT}}_{500}$ for \textit{Planck}. This binning scheme ensures that the first bin has a size exceeding $0.5\,\rm arcmin\rm$ for SPT and $1\,\rm arcmin\rm$ for \textit{Planck}.\\
In our analysis, we use the $95$ and $150\,\rm GHz\rm$ channels of SPT and the $100$, $143$, and $353\,\rm GHz\rm$ channels of \textit{Planck}. The other channels are still used in the component separation process, but are not included in the fit itself. This approach is motivated by the low signal-to-noise ratio of the cluster signals expected at these frequencies. By excluding them, we enhance computational efficiency with minimal loss of information and reduce the risk of residual contamination from dominant dust emission in these channels.\\
% SPT $220\,GHz$ and \textit{Planck} $217\,GHz$, $545,GHz$, and $857\,GHz$
We compare the model with the data using the following likelihood:
\begin{equation}
    \ln{\mathcal{L}}_{SZ}(\theta)\propto \sum_{i}\Bigg(y_i-I_{\text{SZ,i}}\Bigg)^TC_i^{-1}\Bigg(y_i-I_{\text{SZ,i}}\Bigg)\,,
\end{equation}
where $\ln{\mathcal{L}}_{SZ}(\theta)$ is the Gaussian log-likelihood associated with the SZ profiles $y_{i}$ and the associated template $I_{SZ,i}$ obtained integrating the Pressure profile in Equation \ref{pressure} with parameters $\theta=[P_0,c_{500},\alpha,\beta,\gamma]$. Since we reconstruct the SPT CMB using a linear combination of different channels, there is a possibility that noise from one map might inadvertently affect the others during the background subtraction process. To account for this, we compute the two covariance matrices $C_i$, for both SPT and \textit{Planck}. We generate 100 profiles from the cleaned maps in regions where we apply the same cleaning technique, but we do not expect the presence of cluster signals. For both instruments, we concatenate the profiles derived from each map and calculate the covariance among all these profiles. This approach ensures that the covariance accounts for both instrumental noise and correlations between channels. However, it is important to note that we consider SPT and \textit{Planck} channels to be uncorrelated with each other.\\
We employ the Cobaya MCMC \citep{torrado2021cobaya} framework to perform the fit and uncertainty estimation, exploring various combinations of parameters to optimize the convergence of our chains. However, to constrain the shape of the pressure profile from the inner regions to the faint outskirts of the cluster, we allow three parameters to vary freely: the amplidude $P_0$, the inner slope $\gamma$ and the outer slope $\beta$. The other two parameters are held fixed at their universal values \citep{arnaud2010universal}: $c_{500}=1.177$ and $\alpha =1.051$. We employ uniform priors in combination with the likelihood to estimate the posterior distribution. With the best-fit parameters in hand, we use Equation \ref{pressure} to derive the pressure profile $P_{SZ}=P(r)$.
\subsection{Joint X-ray and Millimetric fit}\label{JointFit}
Our goal is to obtain comprehensive thermodynamic profiles and the hydrostatic mass for this high-redshift system, extending our analysis to larger radii. However, as detailed in Section \ref{X-ray-Fit}, constraining the spectroscopic temperature in the outermost regions of the cluster at such high redshift poses a significant challenge. To overcome this limitation, we intend to merge the density profile from the X-ray fitting with the pressure profile derived from the millimetric analysis. This combined approach allows us to extract the remaining thermodynamic profiles, even in these outer regions.\\
Nevertheless, combining information from instruments with distinct resolutions, such as X-ray and SZ observations, can introduce various systematic uncertainties into our analysis. As explained in previous studies \citep{bourdin2017pressure, kozmanyan2019deriving, wan2021measuring, federico2022chex}, although both methods probe the same gas, discrepancies can arise between X-ray-only and SZ-only analyses.
The sources of these differences can be attributed to the geometry of the system, calibration uncertainties of the instruments, or due to the underlying cosmological framework. To account for the possibility of such systematic effects, we introduce a free normalization parameter, denoted $\eta_T$, which scales the X-ray and millimeter inferences of the ICM profiles accordingly.
In the ideal scenario, we would expect $\eta_T=1$, but in more realistic cases $\eta_T$ is expected to deviate from one due to the various factors mentioned earlier. As detailed in \citet{kozmanyan2019deriving}, 
$\eta_T$ has a straightforward dependence on these properties and can be expressed as a product of three terms:
\begin{equation}
    \label{eq:eta_dependence}
    \eta_T= \mathcal{C}\times \mathcal{B} \times b_n \,.
\end{equation}
The first term relates to quantities tied to the cosmological model. X-ray and SZ signals for a given cluster exhibit different dependencies on the cosmic distance from the cluster’s redshift. The X-ray density depends on the ICM emissivity,
thus involves factors of the angular diameter distance, $D_A(z)$, in particular $n_e \propto D_A(z)^{-3/2}$ \citep{mantz2014cosmology}. In contrast, the SZ signal is proportional to the integral along the line of sight of the electron pressure, and thus $P_{\text{SZ}} \propto D_A(z)^{-1}$. Combining the two different dependencies, we get $\mathcal{C} \propto D_A(z)^{1/2}$. Additionally, the X-ray surface brightness is influenced by the chemical composition of the gas, predominantly hydrogen and helium. This dependence in Equation \ref{eq:Xemission} comes from the cooling function $\Lambda(T,Z)$. Assuming negligible contributions from other elements to the X-ray continuum, we can express the contribution of hydrogen and helium separately $\Lambda(T,H,Y) \propto \Lambda_H \times (1+4n_{\text{He}}/n_p)$. This introduces an additional cosmological bias dependent on the value of the helium abundance $Y$, which is added to the cosmological part of the normalization parameter. The total cosmological systematic can be parameterized as $\mathcal{C} \propto D_A(z)^{-1/2} \times Y^{-1/2}$. \\
The second term of Equation \ref{eq:eta_dependence} encapsulates any morphological bias resulting from the geometry of the system. Throughout our analysis, we assume perfect spherical symmetry. Given that the X-ray and SZ signals have different dependencies on density, an eventual asphericity would affect the normalization parameter. Moreover, more realistic gas distributions may include clumpiness, which, like asphericity, would enhance the X-ray emission of the cluster. Taking into account these factors, the geometric part of the normalization parameter can be expressed as $\mathcal{B}\propto C_{\rho}^{1/2} /e_{\text{LOS}}^{1/2}$, where $e_{\text{LOS}}$ and $C_{\rho}$ account for cluster asphericity and clumpiness, respectively. For detailed derivations of the dependence of the cosmological and geometrical systematic, refer to  \citet{kozmanyan2019deriving}. The last term accounts for all other potential systematics due to cross-calibration issues between \textit{Chandra} and XMM-\textit{Newton}, since a variation in temperature for the cluster results in a variation of the normalization parameter in the same direction. Finally, other potential biases may be associated with assumptions and approximations in our methodology, such as profile modeling or the fitting procedure.\\
To conduct our joint X-ray and millimetric fit, we need to account for all the potential systematic effects described above, encapsulated within the normalization parameter. We use the pressure profile from Equation \ref{pressure} to construct both the SZ brightness template (as defined in Equation \ref{ISZ}) and a spectroscopic temperature template derived from the ideal gas law:
\begin{equation}
    \label{etaequation}
    kT(r)=\eta_T \times P(r)/n_e(r)\,.
\end{equation}
Here, we fix $n_e(r)$ to the form obtained in Section \ref{X-ray-Fit}. We fit this template to two projected X-ray temperatures extracted in $[0.0-0.15]R^{\text{SPT}}_{500}$ and $[0.15-0.8]R^{\text{SPT}}_{500}$,  ensuring that we have a signal-to-noise ratio (SNR) that exceeds 30\% in both bins.\\
As detailed in Section \ref{millimetricfit}, we perform the fit using the Cobaya MCMC framework, allowing the parameters $P_0$, $\beta$, $\gamma$ and the normalization parameter $\eta_T$ to vary freely. 
After obtaining the best-fit parameters through joint X-ray and millimetric fit, we employ Equation \ref{pressure} to derive the pressure profile $P_{\text{SZ,X}}=\eta_T P(r)$. This pressure profile provides valuable information about the thermodynamics of the galaxy cluster, enabling us to further explore its properties and derive other relevant profiles.

\section{Results}
\label{Result}
\subsection{Parametric Modeling and Pressure profile}
\begin{figure*}
    \centering
    \includegraphics[width=0.8\linewidth]{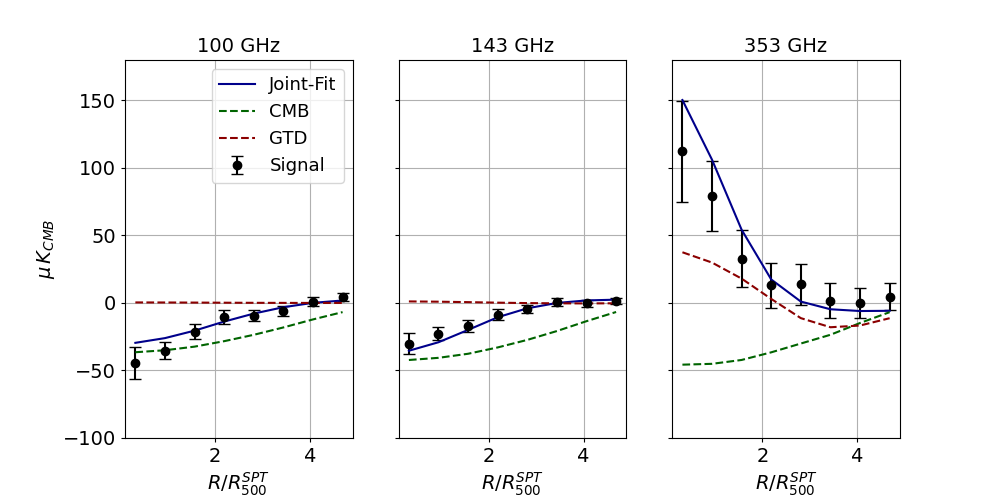}\\
    \includegraphics[width=0.8\linewidth]{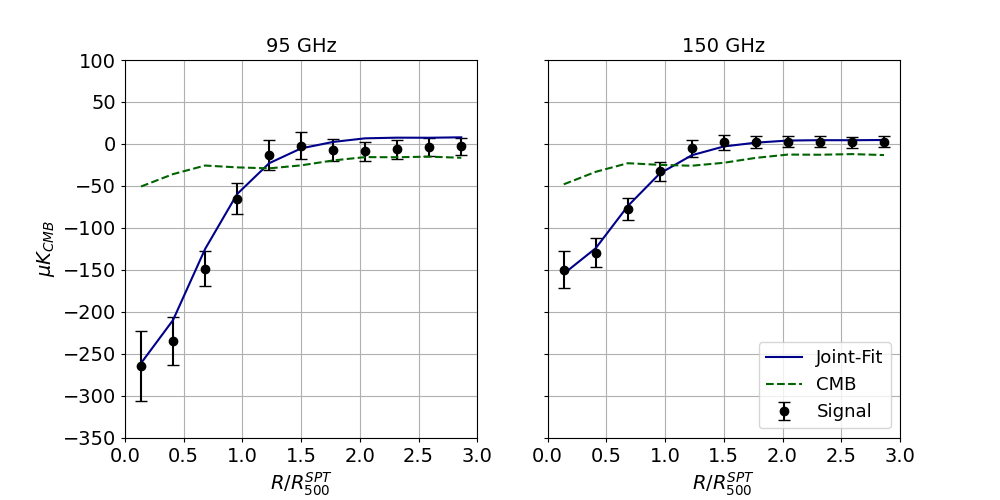}\\
    \caption{Radial profile of the SZ brightness around the X-ray peak of SPT-CLJ0615-5746 of the \textit{Planck} (upper pannels) and SPT (lower pannels) channels used in the fitting procedure. The black dots represent the radial average values on the cleaned maps with their respective errors, and the green and brown lines represent the CMB and GTD respectively. The blue line represents the best fit from the joint X-ray and millimetric analysis (Section \ref{JointFit}).}
    \label{fig:SZ-profile}
\end{figure*}
\begin{figure}
    \centering
    \includegraphics[width=1\linewidth]{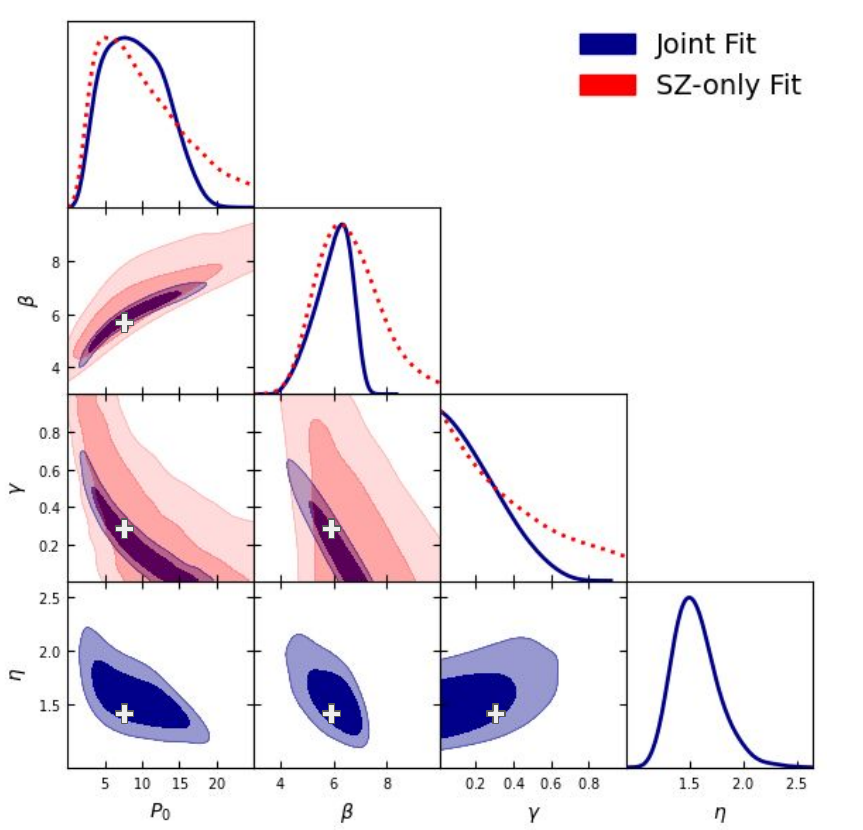}
    \caption{Corner plot illustrating the results of the MCMC analysis for the 
 \citet{nagai2007effects} model parameters and normalization parameter. The joint fit is represented by the blue lines and contours, while the red lines represent the SZ-only fit. Each panel along the diagonal displays the marginalized posterior distribution for a single parameter, while the off-diagonal panels show the two-dimensional projections highlighting the covariances between parameter pairs. The contours represent the 68\% and 95\% confidence intervals. There is no red counterpart for the $\eta_T$ parameter since it is not fitted in the SZ-only Fit. The white crosses represent the position of the best-fit (maximum likelihood) parameters associated with the joint fit.}
    \label{fig:cornerplot}
\end{figure}
Figure \ref{fig:SZ-profile} shows  the outcomes of our joint fit procedures applied to the SZ radial frequency profiles obtained from both the SPT and \textit{Planck} channels used in our analysis. These profiles are presented alongside the background and foreground components. We calculate the best-fit profile (blue lines) using the set of parameters that maximize the likelihood of our chains. As shown in Figure \ref{fig:cornerplot} our joint fit approach, which combines X-ray and millimetric data, improves the constraint on the profile parameters \citet{nagai2007effects}, in particular the upper limit of the internal shape. A lower value of this upper limit also results in a better constraint of the other two parameters. This improvement is primarily attributed to the high spatial resolution of the \textit{Chandra} telescope, allowing better constraints on the shape of the pressure profile within the inner region of the cluster (inside $R^{\text{SPT}}_{500}$).\\

Although the joint-fitting approach clearly enhances our ability to constrain the pressure structure within the inner regions of the cluster, it is worth noting that we obtain a relatively high value for the normalization parameter. In particular, we found that the mean posterior along with the 16th and 84th percentile is $\eta_T=1.46^{+0.15}_{-0.22}$. This value implies that for this particular system, there is a significant difference of around $50\%$ between the ICM properties derived from our X-ray analysis and those obtained through the SZ analysis. Despite the relatively high value obtained, our findings reveal a lower result compared to the value reported in \citet{bourdin2017pressure}. In their study, they performed a joint fit using \textit{Chandra} and Planck-only data sets for this particular cluster, producing a value of $\eta_T\sim1.7$. These outcomes suggest that the inclusion of SPT data in the joint fit effectively reduces the systematics observed between the X-ray and millimeter data sets. This reduction can be attributed to the higher resolution provided by SPT in contrast to Planck, enabling the analysis of pressure structures at radii closer to the X-ray resolution.\\ 
%\section{ICM profiles}
%\label{icmprofiles}
\subsection{Angular sectors}
\begin{figure}
    \includegraphics[width=0.91\linewidth]{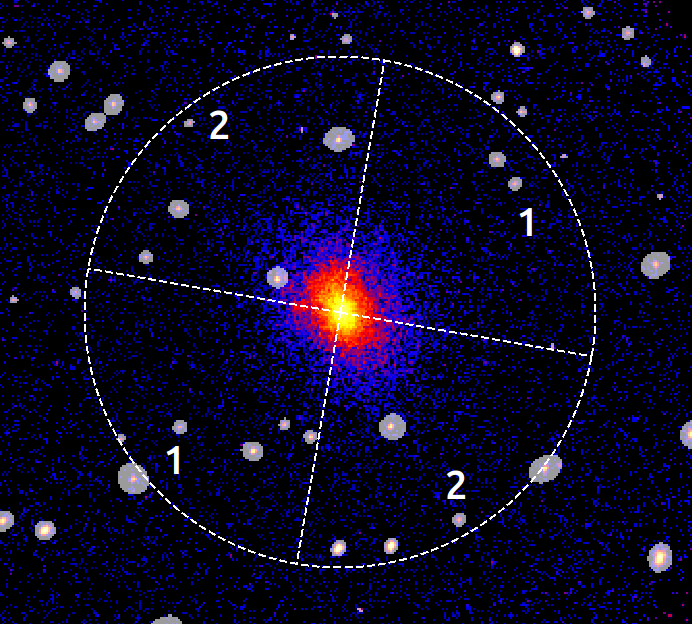}\\  
    \caption{X-ray Chandra image after background subtraction, captured within the energy range of [0.5,2.5] keV.The dashed white circle denotes the position at a radius of $1.4\,R^{\text{SPT}}_{500}$. The white sector outlines the two distinct regions under separate analysis, each identified by a numerical label. The first region (labeled as 1) corresponds to the sector aligned with the minor axis, while the second region (labeled as 2) is oriented along the major axis. The shaded white areas represent the masked regions used to exclude point sources from the analysis. 
    }
    \label{fig:Sector_brightness}
\end{figure}
\begin{figure*}
    \centering
    \includegraphics[width=0.48\linewidth,height=0.4\linewidth]{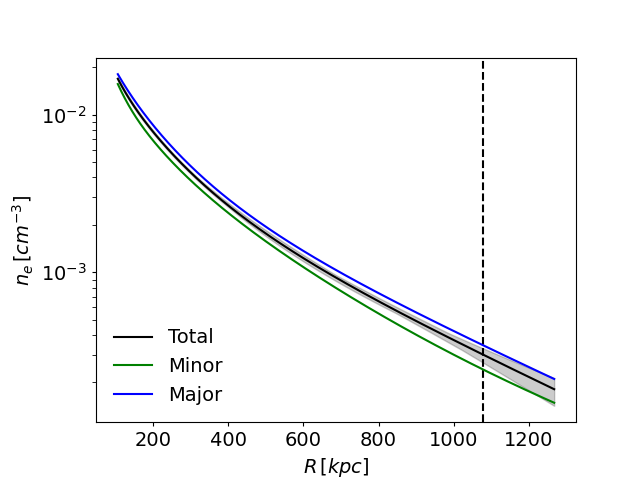}
    \includegraphics[width=0.48\linewidth,height=0.4\linewidth]{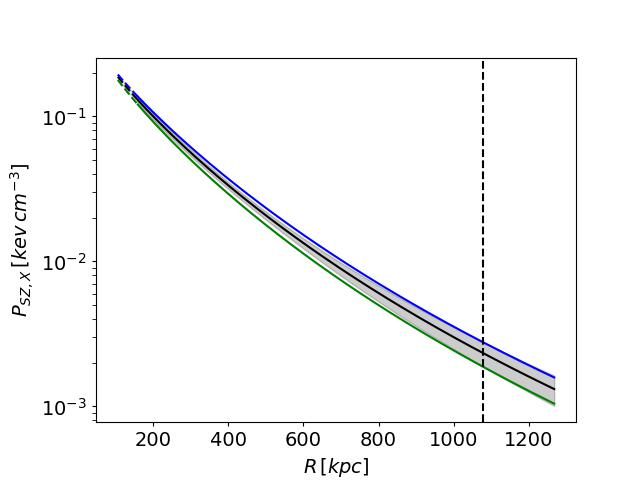}\\
    \includegraphics[width=0.48\linewidth,height=0.4\linewidth]{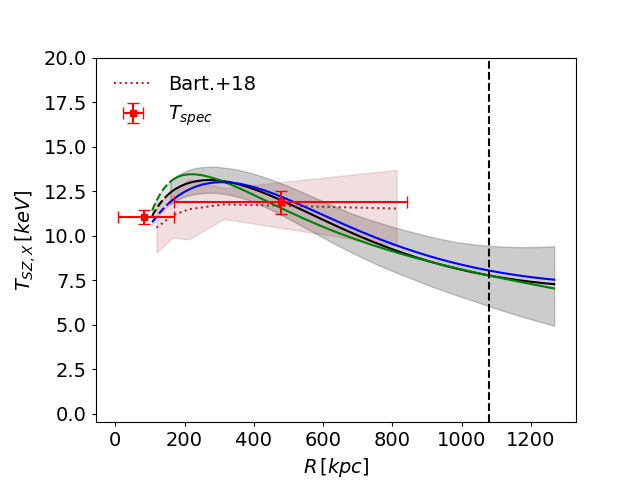}
    \includegraphics[width=0.48\linewidth,height=0.4\linewidth]{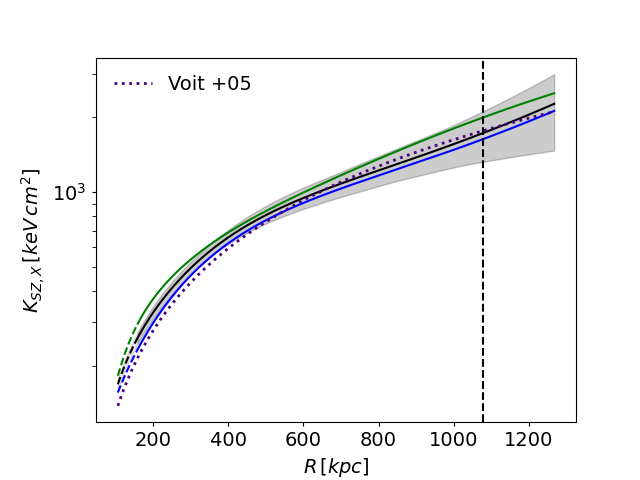}
   \caption{Top: The density and pressure profiles of SPT-CLJ0615-5746 derived independently from the X-ray analysis and the joint fit, respectively. Bottom left: 3D temperature profile derived from the ideal gas law. The red points represent the spectroscopic temperature bins used in the joint fit. The brown dotted line shows the 3D temperature from \citet{bartalucci2017resolving}, derived from a deprojection of the X-ray spectroscopic temperatures of this system. Bottom right: Cluster entropy profile derived from Equation \ref{entropy}. The violet dotted line shows the theoretical entropy profiles from the gravity-only simulations of \citet{voit2005baseline}. In each figure, the black solid line represents the profile derived from the total density profile, while the green and blue lines represent the analysis conducted along the minor and major axes, respectively. The dashed section line marks the region inside the innermost Chandra spectroscopic temperature data point. The vertical dashed line represents the estimation of $R^{\text{HE}}_{500}$. The shaded region corresponds to the 68\% credible interval.} 
    \label{fig:temp_and_entropy}
\end{figure*}
As explained in Section \ref{JointFit}, the $\eta_T$ parameters is influenced primarily by two factors: one associated with the underlying cosmological framework and the other related to the properties of the ICM distribution. To separate the contributions of these two factors, we consider the possibility of the cluster deviating from spherical symmetry. This assumption is motivated by our analysis of the surface brightness map of the cluster, as illustrated in Figure \ref{fig:wavletmap}. This map reveals a departure from perfect sphericity, displaying an elliptical-like shape that is not visible in the SPT and \textit{Planck} maps due to their lower angular resolutions.\\
In this section, we follow the methodology detailed in Section \ref{X-ray-Fit} to derive two separate density profiles for sectors aligned with the cluster's minor and major axes. Figure \ref{fig:Sector_brightness} visually represents the specific regions from which we extract the surface brightness profile. The regions identified by number one are associated with the circular sector along the minor axis, while the remaining regions correspond to the major axis. To facilitate this analysis, we applied masks to the opposing regions, ensuring that only photon counts within each sector were considered.
Subsequently,  the two density profiles are employed in Equation \ref{etaequation} to perform the joint fit for the major and minor axes. We left the SZ analysis unchanged, which means that we do not perform the same division for the SZ profiles. In this way, we introduce an additional geometrical bias that would affect the $\eta_T$ parameter. Our analysis yields $\eta_{\text{T,Minor}}=1.24^{+0.12}_{-0.19}$ and $\eta_{\text{T,Major}}=1.64^{+0.17}_{-0.25}$ for the minor and major axes, respectively. These values are the mean posterior along with the 16th and 84th percentiles. This discrepancy can be attributed to the elliptical shape of the cluster, since at higher radii the density measured along the major axis is higher than the one along the minor axis, with the total one being the mean between the two, as shown in the top left panels of Figure \ref{fig:temp_and_entropy}.  These results confirm that a portion of the discrepancies between the X-ray and SZ analyses can be attributed to geometrical effects specific to this system. However, it is worth noting cross-calibration issues might also contribute to these discrepancies. On average \textit{Chandra} temperatures are $\sim 10-15\%$ higher than XMM-\textit{Netwon}, consequently increasing the $\eta_T$ parameter the same amount. Taking into account these considerations, our values are consistent within the $\sim 15-20\%$  scatter, around the unity, of the $\eta_T$ distribution found in previous work \citep{bourdin2017pressure, kozmanyan2019deriving, wan2021measuring, federico2022chex}.\\
Moreover, in our joint fit the $\eta_T$ parameter comes from Equation \ref{etaequation} that becomes:
\begin{equation}
    \frac{\eta_T}{n_e(r)}= \frac{kT(r)}{P(r)}\,.
\end{equation}
Since we use the same two spectroscopic temperatures bins (red points in Figure \ref{fig:temp_and_entropy}) in the three regions and the SZ analysis is unchanged (i.e., the pressure $P(r)$ does not change), the $\eta_T$ parameter adjusts accordingly to compensate for differences in the density profile. This leads to a lower value for the minor axis and a higher value for the major axis.\\

Toward the end of our fitting procedure, we obtain two radial profiles for both the total and the angular sector analysis : the density profiles from the \textit{Chandra} X-ray fit and pressure profiles from the joint X-ray and millimetric fit, represented as $P_{\text{SZ,X}}(r)= \eta_T P(r)$ (top panels of Figure \ref{fig:temp_and_entropy}). For our analysis, we consider a radial range spanning from approximately $0.15\,R^{\text{SPT}}_{500}$ up to around $\sim 1.2\,R^{\text{SPT}}_{500}$. The upper limit is determined by the emission mesure from \textit{Chandra} analysis, where the signal becomes undetectable on the edges of the cluster, as $\Sigma_X$ scales with $n_e^2$. On the contrary, the lower limit is dictated by the first spectroscopic temperature bin used in the joint fit. Ultimately, we combine these profiles to unravel the thermodynamic properties of the gas of the galaxy cluster and the hydrostatic mass.
\subsection{Temperature and Entropy profiles}
Assuming that the gas behaves as an ideal gas, we can use the equation \ref{etaequation} to recover the 3D temperature profile $kT_{SZ,X}(r)$ and the ICM entropy profile $K_{SZ,X}$ from the following equations:
\begin{equation}
    \label{temperature}
    kT_{\text{SZ,X}}(r) = \eta_T \frac{P(r)}{n_{e}(r)}\,,
\end{equation}
\begin{equation}
    \label{entropy}
    K_{SZ,X}(r) = \eta_T \frac{P(r)}{n_{e}(r)^{5/3}}\,.
\end{equation}
By combining the two datasets, we are able to determine the 3D temperature and entropy profiles up to radial distances beyond $R^{\text{HE}}_{500}$ without the need for extrapolating a parametric temperature profile in regions where the spectroscopic temperature is not well constrained.\\
The profiles obtained are presented in the lower panels of Figure \ref{fig:temp_and_entropy}. Remarkably, we observe a high level of agreement with the 3D temperatures derived from an X-ray-only analysis, as reported in \citet{bartalucci2017resolving}, within the range covered by their radii. These findings emphasize the consistency and robustness of our joint fit method, enabling the extraction of thermodynamic profiles up to higher radii in comparison to an X-ray-only analysis.\\
Moreover, also our entropy profiles are consistent with the expected profile of \citet{voit2005baseline}. Here, we highlight that our method is mostly limited by our ability to extract a density profile regardless of the availability of spectroscopic temperature measurements.\\
Alongside the total profiles (black lines) we show the results obtained from our angular sector-based analysis (green and blue lines). The temperature profiles are computed using Equation \ref{temperature}, along with the density profiles extracted in the three regions. Since our joint fit leads to a constant ratio $\eta_T/n_e(r)$, we get the same 3D temperature profiles regardless of where we extract the emission measure data to fit the density profile. This does not apply for the entropy profiles. From Equation \ref{entropy} it is clear that density plays a more dominant role compared to pressure, leading to differences (within the errors) between our sector-based profiles on the outskirts of the cluster.
\subsection{Hydrostatic mass profile}
\begin{figure}
    \centering
    \includegraphics[width=1\linewidth]{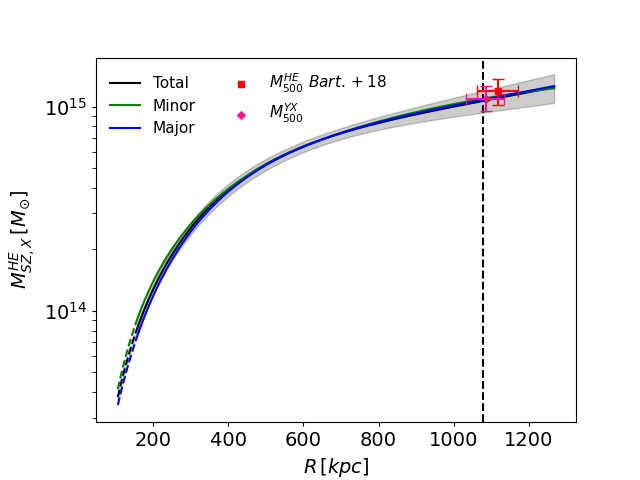}
    \caption{The mass profiles of SPT-CLJ0615-5746, derived under the assumption of hydrostatic equilibrium using Equation \ref{Mass_SZX}. The black solid line represents the mass profile derived from the total density profile, while the green and blue lines represent the analysis conducted along the minor and major axes, respectively. The dashed section line marks the region inside the innermost Chandra spectroscopic temperature data point. The vertical dashed line represents the estimate $R^{\text{HE}}_{500}$. The shaded region corresponds to the 68\% credible interval. The purple point is our $M^{\text{YX}}_{500}$ estimated using the $Y_X-M$ relation, while the red point is $M^{\text{HE}}_{500}$ from the X-ray only analysis of \citet{bartalucci2018resolving}. }
    \label{fig:Mass_profile}
\end{figure}
Under the assumption of  spherical symmetry and  hydrostatic equilibrium, the integrated mass profile of a galaxy cluster can be expressed as:
\begin{equation}
 \label{Mass_T}
    M^{HE}(\leq R) = -\frac{r^2}{G\mu m_p n_e(r)}\frac{d P_{e}(r)}{dr}\, ,
\end{equation}
where $\mu=0.596$ is the gas mean molecular weight, $G$ is the Newton's constant and $m_p$ is the proton mass. We extract the hydrostatic mass profile by combining the pressure profile measured from our joint fit $P_{SZ,X}=\eta_T P(r)$ and the density profile $n_e(r)$ measured independently from the X-ray data. Combining the two profiles, we get:
\begin{equation}
    \label{Mass_SZX}
    M^{HE}_{\text{SZ,X}}(\leq R) = - \eta_T \frac{r^2}{G\mu m_p n_e(r)}\frac{d P(r)}{dr}\,.
\end{equation}
Given that we can constrain the pressure profile from the SZ signal to distances greater than $2\,R_{500}$, the upper limit of our hydrostatic mass estimation is determined by the radial range covered by the density profile.
Figure \ref{fig:Mass_profile} displays the results of our analysis, showing the mass profiles derived from our calculations. Using the critical density of the Universe at the cluster's redshift, we calculate the over-density contrast by integrating the mass profile. This enables the determination of $R^{\text{HE}}_{500} = 1077.61^{+69.30}_{-56.77}\,\rm kpc\rm$ (vertical dashed line of Figure \ref{fig:Mass_profile}), from which we subsequently compute $M^{\text{HE}}_{500} = 10.67^{+0.62}_{-0.50}\,\, 10^{14}\,M_{\odot}$.\\

We also present hydrostatic mass estimations for our analysis based on the angular sector. Profiles are calculated using Equation \ref{Mass_SZX}, which incorporates the density profiles extracted from the three regions. Similarly to 3D temperature profiles, the constant ratio $\eta_T/n_e(r)$ ensures consistent hydrostatic mass measurements regardless of the region considered.\\
Our findings indicate that the stability and consistency of our mass and temperature estimates persist, regardless of the chosen region for extracting emission measure data to fit the density profile. This implies that variations in the normalization parameter, whether higher or lower, effectively compensate for any differences in the density profile.\\

We compare the hydrostatic mass estimates obtained from our joint fit with other mass estimates derived from independent studies or methods. To make a meaningful comparison, we interpolate our mass profiles at the radii corresponding to the locations of the other mass estimates. Despite the elongated shape of this system, we find good agreement with the integrated mass estimation based on the $M_{500}-Y_{SZ}$ and $M_{500}-\xi$ scaling relations. In particular, our result aligns with the $\xi$-mass scaling relation used in the SPT catalog \citep{bleem2015galaxy}. We find $M^{\text{HE}}_{\text{SZ,X}}/M^{\text{SPT}}_{500}=1.03 \pm 0.17$ ($M^{\text{HE}}_{\text{SZ,X}}$ is interpolated at $R^{\text{SPT}}_{500} = 1059 \,\rm kpc\rm$). Furthermore, we compare our hydrostatic mass estimate with the integrated mass estimate of the $Y_X$ proxy.  We calculate the value of $M^{Y_{X}}_{500} = 10.91_{-1.42}^{+1.70}\, 10^{14}M_{\odot}$ and $R^{Y_{X}}_{500}= 1085.00_{-49.52}^{+53.67}\, \rm kpc\rm$ iteratively using the $M_{500}-Y_{X}$ scaling relation calibrated with estimates from Chandra data \citep{vikhlinin2009chandra}. Here $Y_X$ is defined as the product of the gas mass computed at $R^{Y_{X}}_{500}$ and the temperature measured in $[0.15-1.0]\,R^{Y_{X}}_{500}$. Finally, we compare our two estimates by interpolating the hydrostatic mass at $R^{Y_{X}}_{500}$. In this case, also, we find excellent agreement with $M^{\text{HE}}_{\text{SZ,X}}/M^{Y_{X}}_{500}= 0.98 \pm 0.18$. \\
Finally, we compare our results with the X-ray only hydrostatic mass published in \citet{bartalucci2018resolving} who determined $R^{\text{HE,B}}_{500}=1119^{+52}_{-58} \,\rm kpc\rm$ for this cluster, combining XMM-\textit{Newton} and \textit{Chandra} observations.
In particular, we find $M^{\text{HE}}_{\text{SZ,X}}/M^{\text{HE,B}}_{500}= 0.92 \pm 0.18$, where $M^{\text{HE}}_{\text{SZ,X}}$ is interpolate at $R^{\text{HE,B}}_{500}$.
An X-ray only estimates of $M^{\text{HE}}_{500}$ of such high redshift cluster are possible thanks to the deep \textit{Chandra} observations available for this system, since it relies on the radial coverage provided by temperature data. The consistency between our mass and an X-ray only mass estimates provides confidence in extending trust to our results at larger radii, where we do not have enough statistics to constrain the spectroscopic temperatures. These findings suggest the potential extension of our analysis to clusters where we do not have such deep observation, encouraging us to apply our method even to higher-redshift clusters.

\section{Summary and Conclusions}
\label{Summaryandconclusion}
In this paper, we have introduced a multiwavelength technique to resolve the gas properties of the most distant galaxy cluster detected by \textit{Planck}, located at
$z=0.972$. Our approach combines data from \textit{Planck} and the South Pole Telescope (SPT) with \textit{Chandra} observations, allowing us to extract individual pressure and density profiles for this cluster up to $R>R_{500}$. We employ two distinct fitting methods. The first involves independent analyses of the data from \textit{ Chandra} and SPT-\textit{Planck} (SZ-only) data. In contrast, the second method employs a joint fit that combines the multiwavelength data from all three instruments. For the latter, we introduce a free normalization parameter $\eta_T$ to build a temperature template. This template is then fitted exclusively in the innermost region of the cluster $R < 0.8\, R_{500}$, where there are enough photons count to constrain the spectroscopic temperatures. Our main findings and results regard our fitting methods are summarized below.
\begin{itemize}
\item Our multiwavelenght approach allows us to constrain  the thermodynamic profiles across a wider range of spacial scales compared to a SZ-only or X-ray-only analysis.  The high resolution provided by \textit{Chandra} facilitates a more precise constraint on the shape of the pressure profile within the inner region of the cluster. Meanwhile, the synergy between SPT and \textit{Planck} allows us to extend these constraints to larger spatial scales. The reduced uncertainty in the shape of the profile translates into improved accuracy in determining the best-fit parameters of the \citet{nagai2007effects} profile.

\item Our analysis reveals a high value for the normalization parameter, specifically $\eta_T= 1.46^{+0.15}_{-0.22}$. We observe a significant deviation from the ideal case ($\eta_T=1$) , and this holds whether we extract the density in concentric rings or in sectors along the principal axes. Even when using the latter approach, which helps mitigate geometric effects due to the departure of this system from spherical symmetry, we still obtain $\eta_T>1$. The discrepancy in the values of $\eta_T$ suggests that a straightforward combination of \textit{Chandra} X-ray and millimetric data, without accounting for systematic differences, is not feasible. This highlights the importance of incorporating a normalization parameter in our analysis, as it encodes these systematic comprehensively.
\end{itemize}
We then combine the resulting profile from our fitting procedure to calculate the hydrostatic mass, 3D temperature, and entropy profiles up to $R>R_{500}$ without the need for precise spectroscopic temperature measurements in the outskirts of the cluster. Here the main results:
\begin{itemize}
    
    \item  Our multiwavelength approach allows us to extend our coverage to a radial range primarily limited by our ability to extract a density profile rather than by the availability of spectroscopic temperature measurements. This highlights the potential of our method to provide insights into thermodynamic profiles and hydrostatic mass of galaxy clusters at large radii, even in cases where obtaining accurate spectroscopic temperature data is challenging.
    \item Our 3D temperature profile aligns with the X-ray-only analysis conducted in \citet{bartalucci2017resolving}, specifically within the range covered by their radii. This consistency ensures the robustness of our analysis, providing confidence in extending trust to the results at larger scales.
    \item We find that the hydrostatic mass and temperature estimates remain stable and consistent, regardless of the region-specific density profiles used. The variation in the normalization parameter effectively compensates for the differences between the X-ray and millimetric selected regions of the cluster, resulting in similar mass estimate. 
    \item We compare our value of $M^{\text{HE}}_{SZ,X}$ with other estimates. In particular, we find a good alignment with the $\xi$-mass scaling relation used in the SPT catalog \citep{bleem2015galaxy} and with the $M_{500}-Y_X$ scaling relation \citep{vikhlinin2009chandra}. We also find consistency with the hydrostatic mass estimates based only on X-rays from \citet{bartalucci2018resolving}.
\end{itemize}
This study represents the initial comprehensive application of the methodology developed in \citet{bourdin2017pressure} and \citet{oppizzi2023chex} to investigate and resolve the ICM properties of high-redshift objects. We successfully combined data from three different instruments, allowing us to analyze a broad radial range of the cluster. \textit{Chandra} and SPT provide insights into the cluster structure up to the inner regions (approximately $0.15\,R_{500}$), while \textit{Planck} data serve as a robust anchor at larger angular scales and provide improved estimates of the cosmic microwave background and the galactic thermal dust emissions. \\
The introduction of a normalization parameter offers several advantages. Firstly, it exploits the synergy between X-ray and mm observations by resolving cluster cores close to an arc-second angular resolution while probing cluster peripheries out to intra-cluster radii barely accessible to X-ray spectroscopy alone.  Moreover, it highlights the systematic uncertainties arising from the combination of different instruments.
Our work demonstrates the power of combining multiwavelength observations to obtain a comprehensive understanding of galaxy clusters. This technique opens up new possibilities for studying the gas properties and mass distribution in clusters even when spectroscopic temperature measurements are challenging to obtain, in particular, for high redshift clusters.
For distant galaxy clusters of SZ cluster catalogs also observed in X-rays, these results motivate future studies related to the evolution with cosmic times of the hot gas thermodynamics and the shape of matter haloes.

\begin{acknowledgements}
    CM, FO, HB, PM, and FDL acknowledge financial contribution from the contracts ASI-INAF Athena 2019-27-HH.0, "Attività di Studio per la comunità scientifica di Astrofisica delle Alte Energie e Fisica Astroparticellare" (Accordo Attuativo ASI-INAF n. 2017-14- H.0), from the European Union’s Horizon 2020 Programme under the AHEAD2020 project (grant agreement n. 871158), support from INFN through the InDark initiative, from “Tor Vergata” Grant “SUPERMASSIVE-Progetti Ricerca Scientifica di Ateneo 2021”, and from Fondazione ICSC , Spoke 3 Astrophysics and Cosmos Observations. National Recovery and Resilience Plan (Piano Nazionale di Ripresa e Resilienza, PNRR) Project ID CN00000013 ‘Italian Research Center on High-Performance Computing, Big Data and Quantum Computing’ funded by MUR Missione 4 Componente 2 Investimento 1.4: Potenziamento strutture di ricerca e creazione di "campioni nazionali di R\&S (M4C2-19 )" - Next Generation EU (NGEU). We acknowledge the financial contribution from the contract Prin-MUR 2022 supported by Next Generation EU (n.20227RNLY3 {\it The concordance cosmological model: stress-tests with galaxy clusters}).
\end{acknowledgements}

\bibliographystyle{aa}
\bibliography{example} % if your bibtex file is called example.bib
\begin{comment}

\end{comment}
\end{document}